\documentclass[preprint,prd,aps,tighten,nofootinbib,amssymb]{revtex4}

\usepackage{graphicx}
\usepackage{epsfig,amssymb,amsmath}

\pdfoutput=1
\usepackage{subfigure}
\usepackage[colorlinks=true,linkcolor=blue,citecolor=blue]{hyperref}
\usepackage{float}

\makeatletter
\def\Hline{%
\noalign{\ifnum0=`}\fi\hrule \@height 2pt \futurelet
\reserved@a\@xhline}
\makeatother

\newcommand{\beq}{\begin{equation}}
\newcommand{\eeq}{\end{equation}}
\newcommand{\bea}{\begin{eqnarray}}
\newcommand{\eea}{\end{eqnarray}}
\newcommand{\bear}{\begin{array}}
\newcommand {\eear}{\end{array}}
\newcommand{\bef}{\begin{figure}}
\newcommand {\eef}{\end{figure}}
\newcommand{\bec}{\begin{center}}
\newcommand {\eec}{\end{center}}

\newcommand{\la}{\left\langle}
\newcommand{\ra}{\right\rangle}
\newcommand{\ds}{\displaystyle}

\def\GEV#1{10^{#1}{\rm\,GeV}}

\def\lrfp#1#2#3{ \left(\frac{#1}{#2} \right)^{#3}}

\begin{document}
\draft
\tighten
\preprint{TU-985,~IPMU14-0334}
\title{\large \bf
Resonant conversions of QCD axions into hidden axions and
suppressed isocurvature perturbations
}
\author{
    Naoya Kitajima\,$^{a}$\footnote{email: kitajima@tuhep.phys.tohoku.ac.jp},
    Fuminobu Takahashi\,$^{a,b}$\footnote{email: fumi@tuhep.phys.tohoku.ac.jp}
}
\affiliation{
$^a$ Department of Physics, Tohoku University, Sendai 980-8578, Japan\\
$^b$ Kavli IPMU, TODIAS, University of Tokyo, Kashiwa 277-8583, Japan
}

\vspace{2cm}

\begin{abstract}
We study in detail MSW-like resonant conversions of QCD axions into hidden axions,
including cases where the adiabaticity condition is only marginally satisfied, and where anharmonic effects are non-negligible. 
When the resonant conversion is efficient,  the QCD axion abundance is suppressed by the hidden and QCD axion mass ratio. 
We find that, when the resonant conversion is incomplete due to a weak violation of the adiabaticity, 
the CDM isocurvature perturbations can be significantly suppressed, while non-Gaussianity of the isocurvature 
perturbations generically remain unsuppressed. 
The isocurvature bounds on the inflation scale can therefore be 
relaxed by the partial resonant conversion of the QCD axions into hidden axions. 
\end{abstract}

\maketitle

\section{Introduction}
\label{sec:intro}
The strong CP problem in the standard model is one of the most profound mysteries
in particle physics, and a plausible solution is the Peccei-Quinn (PQ) mechanism~\cite{Peccei:1977hh}. 
When a global  U(1)$_{\rm PQ}$ symmetry is spontaneously broken, there appears an associated
Nambu-Goldstone (NG) boson ``axion" which is assumed to acquire a tiny mass predominantly from the 
QCD anomaly~\cite{Weinberg:1977ma,Wilczek:1977pj,QCD-axion}.
As a result, the axion is stabilized at the CP-conserving minimum, and the strong CP problem is
dynamically solved. 

In the early Universe, the QCD axion remains massless  until the cosmic temperature
drops down to the QCD scale, $\Lambda_{\rm QCD} \simeq 400$\,MeV. During the QCD phase transition, 
the QCD axion gradually acquires a mass, and it starts to oscillate about the CP conserving minimum
when the mass becomes comparable to the Hubble parameter.
The PQ mechanism is therefore necessarily accompanied by coherent oscillations of the QCD axions,
which
behave as cold dark matter (CDM). The abundance of axion coherent oscillations is given by~\cite{Bae:2008ue}
\beq
	\Omega_a h^2 \simeq 0.195 \theta_i^2 f(\theta_i) \bigg(\frac{F_a}{10^{12}~{\rm GeV}} \bigg)^{1.184},
	\label{eq:Omega_ah2}
\eeq
where $\theta_i$ is the initial misalignment angle of the QCD axion, $F_a$ is the QCD axion decay constant, 
and $f(\theta_i)$ is the anharmonic correction that is a monotonically increasing function of $\theta_i$:  
$f(\theta_i) \sim 1$ for $\theta_i \lesssim 1$ and it grows rapidly as $\theta_i$ approaches $\pi$~\cite{Visinelli:2009zm}.

There are  two possible cosmological problems of the QCD axion dark matter.
One is the overabundance; if the decay constant is of order the GUT or string scale, the axion abundance 
would be many orders of magnitude larger than the observed dark matter unless the initial misalignment
angle is less than $10^{-2}$.  The other  is the tight isocurvature constraint on the inflation scale; if the axion is present
during inflation, it acquires quantum fluctuations of order the Hubble parameter, which induce the CDM isocurvature
perturbations. The upper bound on the inflation scale is so tight that it excludes large-field inflation
 models~\cite{Higaki:2014ooa,Marsh:2014qoa,Visinelli:2014twa}.\footnote{ 
There have been proposed various ways to suppress the axion CDM isocurvautre 
perturbations~\cite{Linde:1990yj,Lyth:1992tx,Dine:2004cq,Jeong:2013xta,Higaki:2014ooa,
Linde:1991km,Kawasaki:2013iha,Folkerts:2013tua,Fairbairn:2014zta}.
}

The QCD axion may not be the only pseudo NG boson in nature; there may be
many axions or axion-like particles (ALPs)~\cite{Masso:1995tw,Masso:2004cv,Jaeckel:2006xm,
Cadamuro:2011fd,Arias:2012az}. Indeed, in a certain class of the compactification
of the extra dimensions, there remain many light axions, some of which may play
an important cosmological role~\cite{Cicoli:2012sz}. For instance, multiple axions in cosmological contexts have
been studied in the so-called axiverse scenario~\cite{Arvanitaki:2009fg,Acharya:2010zx} and the axion 
landscape~\cite{Higaki:2014pja,Higaki:2014mwa}. Suppose that there is another
axion which has a mixing with the QCD axion. Then, as the QCD axion gradually acquires a mass during the 
QCD phase transition, the MSW-like conversions could take place.
Such resonant conversion was studied in Ref.~\cite{Hill:1988bu} 
assuming that adiabaticity condition is satisfied and anharmonic effects are negligible. 

In this letter
we study in detail the resonant conversions of axions, including cases where
the adiabaticity condition is only marginally satisfied or weakly violated, and where the anharmonic
effects are non-negligible. We show that the axion abundance can be suppressed by the mass
ratio of the hidden and QCD axions. This is because it is the number of the axions in a comoving volume that is conserved
during the resonant conversion process. The authors of  Ref.~\cite{Hill:1988bu} claimed that the QCD axion
abundance is suppressed by the square of the mass ratio because the oscillation amplitude is 
conserved in the conversion process, which however was not confirmed in our analysis.
We shall also study how the isocurvature perturbations are modified
when the resonant conversions take place. Interestingly, we find that, when the resonant conversion 
is incomplete due to weak violation of the adiabaticity condition,  the power spectrum of the 
isocurvature perturbations can be significantly suppressed for certain parameters.
This is because the conversion rate also depends on the initial misalignment angle, and the produced hidden axions
can compensate the isocurvature perturbations of the QCD axions.
Therefore,  the isocurvature constraint on the inflation scale can be relaxed by the incomplete resonant conversions.
On the other hand, non-Gaussianity of isocurvature perturbations are generically non-vanishing even in this case.

The rest of this letter is organized as follows.  In Sec.~\ref{sec:dynamics}, we give our model of axions and study
the dynamics of the axion oscillations, focusing on the resonant conversion processes. 
In Sec.~\ref{sec:resonance}, we show how the axion abundance and isocurvature perturbations are modified by the
resonant conversion.  Sec.~\ref{sec:conc} is devoted to conclusions and discussions.

\section{Set-up}
\label{sec:dynamics}

In this section we  give a model of the QCD and hidden axions which have a non-zero mixing. In the next section we
study the dynamics of axion coherent oscillations, focusing on the resonant conversion between these
two axions.

Let us introduce two complex scalar fields $\Phi$ and $\Phi_H$ with the following interactions
with heavy quarks, 
\beq
	\mathcal{L} = \kappa \Phi Q \bar{Q} + \frac{\lambda}{M_P} \Phi \Phi_H Q_H \bar{Q}_H
\eeq
where $Q$ and $\bar{Q}$ belong to ${\bf 3}+{\bf \bar{3}}$ of SU(3)$_c$,   $Q_H$ and
${\bar Q}_H$ belong to (anti-)fundamental representation ${\bf N}+{\bf \bar N}$ of a hidden gauge symmetry SU(N)$_H$,
and $\kappa$ and $\lambda$ are dimensionless coupling constants. $M_P \simeq 2.4\times \GEV{18}$ is the reduced Planck mass.
This is an extension of the KSVZ hadronic axion model with additional hidden scalar and quarks.  We assume that there are two global U(1) symmetries,
U(1)$_{\rm PQ}$ and U(1)$_H$, which are spontaneously broken by vacuum expectation values of
$\Phi$ and $\Phi_H$. See Table~\ref{tab:charge} for the charge assignment of each field.
The QCD axion $(a)$ and the hidden axion $(a_H)$ appear as (pseudo) NG bosons associated with 
the spontaneous symmetry breaking, and they reside in the phase component of $\Phi$ and $\Phi_H$, respectively. 
We assume that the hidden gauge symmetry SU(N)$_H$ becomes strong and induces a potential for axions 
in the low energy.  If the hidden gauge sector is not thermalized by the inflaton decay, the axion potential is generated when the 
Hubble parameter becomes comparable to the dynamical scale. 
 
The low-energy effective Lagrangian for axions below the dynamical scale of SU(N)$_H$ is given by 
\beq
	\mathcal{L} = \frac{1}{2}\partial^\mu a \partial_\mu a + \frac{1}{2} \partial^\mu a_H \partial_\mu a_H - V(a,a_H) 
\eeq
with the potential, 
\beq
	V(a,a_H) = m_a^2(T) F_a^2 \left[ 1-\cos\bigg(\frac{a}{F_a} \bigg) \right] 
	+ m_H^2 F_H^2 \left[ 1-\cos\bigg( \frac{a_H}{F_H} + \frac{a}{F_a} \bigg) \right],
	\label{Va}
\eeq
where $m_a(T)$ and $m_H$ are the mass of the QCD and hidden axions respectively, 
$F_a$ and $F_H$ are the decay constants of $a$ and $a_H$, and they are comparable to
the corresponding U(1) symmetry breaking scales.\footnote{
It is possible to generalize the potential as
\beq
	V(a,a_H) = m_a^2(T) F_a^2 \left[ 1-\cos\bigg( n_1\frac{a_H}{F_H} + n_2\frac{a}{F_a} \bigg) \right] 
	+ m_H^2 F_H^2 \left[ 1-\cos\bigg( n_3\frac{a_H}{F_H} + n_4\frac{a}{F_a} \bigg) \right],
\eeq
with $n_1$ -- $n_4$ being some integers, for appropriate charge assignments. In the text we focus on the case
of $n_1=0$ and $n_2 = n_3 = n_4 = 1$, but we can straightforwardly extend our analysis to more general cases. 
}%
Here we have shifted $a$ and $a_H$ so that the origins of $a$ and $a_H$ coincide with the potential minimum. 
Then, the equations of motion for the axions are given by
\beq
	\begin{split}
		&\ddot{a} + 3H\dot{a} + \frac{m_H^2 F_H^2 }{F_a} \sin \left( \frac{a_H}{F_H} + \frac{a}{F_a} \right)
		+m_a^2(T) F_a  \sin \left( \frac{a}{F_a} \right)  = 0 \\[2mm]
		&\ddot{a}_H + 3H\dot{a}_H + m_H^2 F_H  \sin \left( \frac{a_H}{F_H} + \frac{a}{F_a} \right) = 0.
	\end{split}
	\label{eq:eom}
\eeq
\begin{table}[tb]
\begin{center} {\tabcolsep = 2mm
	\begin{tabular}{c|cccccc} \hline
		\rule[0mm]{0mm}{0mm} & $\Phi$ & $\Phi_H$ & $Q$ & $\bar{Q}$ & $Q_H$ & $\bar{Q}_H$ \\ \hline
		${\rm U(1)_{PQ}}$ & 1 & 0 & $-$1 & 0 & $-$1 & 0 \\
		${\rm U(1)_{H}}$ & 0 & 1 & 0 & 0 & $-$1& 0 \\ \hline
	\end{tabular} }
\end{center}
\caption{${\rm U(1)_{PQ}}$ and ${\rm U(1)_H}$ charge assignment.}
\label{tab:charge}
\end{table}
If the axions are initially located in the vicinity of the potential minimum, or if the oscillation amplitudes become
much smaller than the corresponding decay constant, the equations of motion can be approximately linearized as
\beq
	\ddot{A} + 3H \dot{A} + M^2 A \approx 0
\eeq
where $A$ and $M^2$ are respectively the column vector of two axion fields and the (squared) mass matrix,
\beq
	A = \begin{pmatrix}a \\ a_H \end{pmatrix} ~~~\text{and}~~~ 
	M^2 = \begin{pmatrix} m_a^2(T) + \big(\frac{F_H}{F_a} \big)^2 m_H^2 && \mu^2 \\ \mu^2 && m_H^2 \end{pmatrix},
\eeq
and we define $\mu^2 = (F_H/F_a) m_H^2$. One can diagonalize the mass matrix by an orthogonal matrix $O$
as
\beq
	\begin{pmatrix} m_1^2 && 0 \\ 0 && m_2^2 \end{pmatrix} = O^T M^2 O ~~\text{and}~~
	\begin{pmatrix} a_1 \\ a_2 \end{pmatrix} = O^T A
\eeq
with $m_2 > m_1$. In our convention, the mass of $a_2$ is always heavier than or equal to that of $a_1$. 
Alternatively, one may define the mass eigenstates by
\beq
	\begin{pmatrix} a'_1 \\ a'_2 \end{pmatrix} = \begin{pmatrix} \cos\alpha && \sin\alpha \\ -\sin\alpha && \cos\alpha \end{pmatrix}
	\begin{pmatrix} a \\ a_H \end{pmatrix}
\eeq
with the mixing angle
\beq
	\tan 2\alpha = -\frac{2\mu^2}{m_a^2(T)+ [(F_H/F_a)^2 - 1] m_H^2}.
\eeq
Throughout this letter we use the previous notation of the 
mass eigenstates $(a_1, a_2)$ with $m_2 > m_1$  which is more convenient when the resonant conversion takes place.

\section{Cosmology of axion resonant conversion}
\label{sec:resonance}

\subsection{Cross-over of mass eigenvalues}
\label{subsec:crossover}
\begin{figure}[t]
\centering
\includegraphics [width = 8cm, clip]{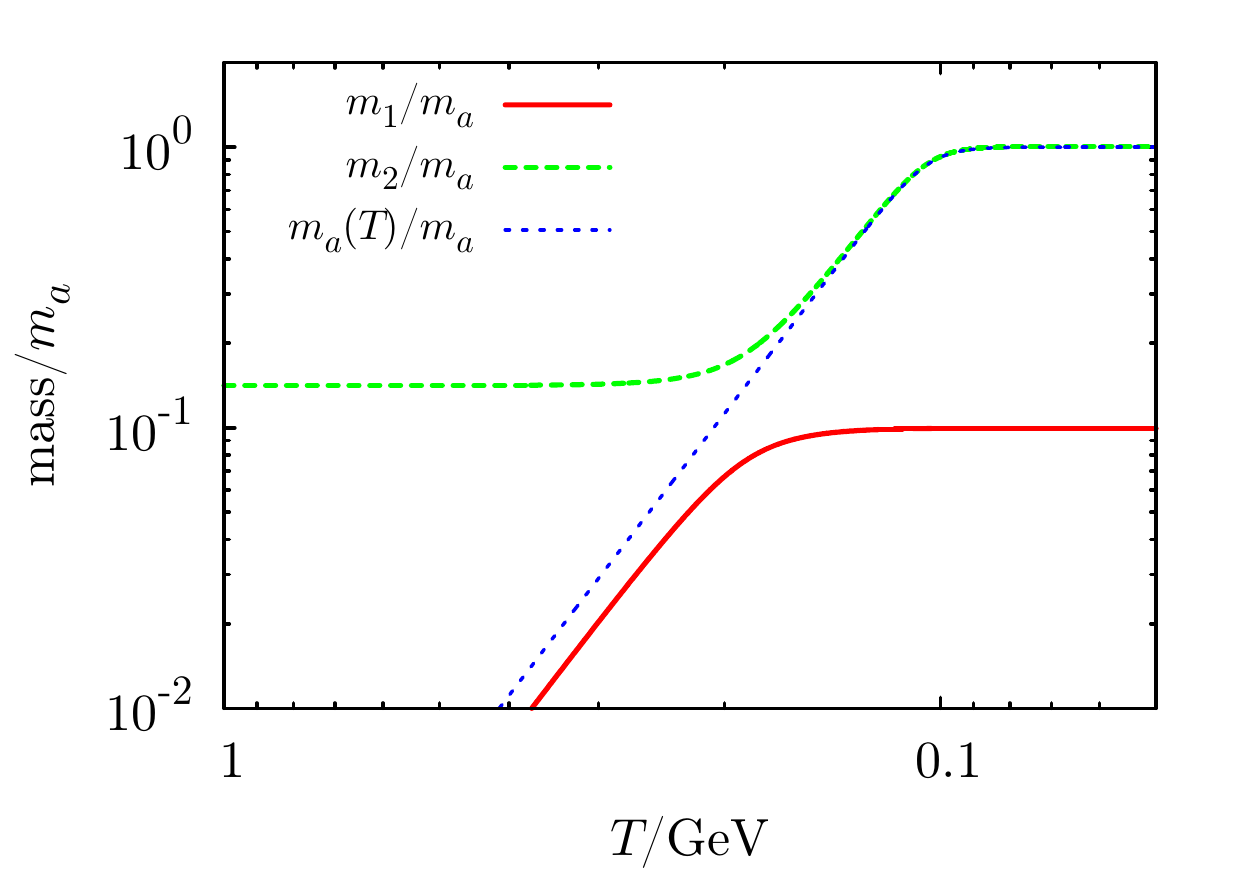}
\caption{
	Evolution of mass eigenvalues as a function of temperature $T$. The masses are
	normalized by the zero-temperature QCD axion mass $m_a = m_a(T=0)$, and the temperature is in the GeV units.
	The solid (red) and dashed (green) lines represent  evolution of the light ($m_1$) and heavy  ($m_2$) mass eigenvalues, respectively, while
	the dotted (blue) line represents the temperature-dependent QCD axion mass, $m_a(T)$.
	We have taken $F_H = F_a = 10^{14}~{\rm GeV}$ and $m_H = 0.1\,m_a$.
}
\label{fig:mass_evolve}
\end{figure}

The QCD axion remains almost massless at temperatures much higher than
the QCD dynamical scale $\Lambda_{\rm QCD} \simeq 400$\,MeV, and it gradually acquires a mass and starts to oscillate during the
QCD phase transition.
The temperature-dependent QCD axion mass is given by \cite{Wantz:2009it}\footnote{
Precisely speaking, $m_a(T)$ parametrizes the potential height in Eq.~(\ref{Va}), and 
the actual mass eigenvalue is slightly different  due to the mixing with the hidden axion. 
}
\beq
	m_a (T) \approx
	\begin{cases} 
		\ds{4.05 \times 10^{-4} \frac{\Lambda_{\rm QCD}^2}{F_a} \left( \frac{T}{\Lambda_{\rm QCD}} \right)^{-3.34} }&~~\text{for}~~T > 0.26 \Lambda_{\rm QCD} \\[1mm]
		\ds{3.82 \times 10^{-2} \frac{\Lambda_{\rm QCD}^2}{F_a}} &~~\text{for}~~ T < 0.26 \Lambda_{\rm QCD}
	\end{cases},
\eeq
where the typical time scale over which the QCD axion mass grows is  the Hubble time $H^{-1}$. 
Therefore, if the zero-temperature mass of the QCD axion, $m_a \equiv m_a(T=0)$,  is heavier than the hidden axion mass, 
there is necessarily a cross-over of the mass eigenvalues. The resonance temperature 
$T_{\rm res}$ is given by the temperature at which the QCD axion mass becomes equal to the hidden axion mass:
\beq
	T_{\rm res} \simeq 0.1 \bigg( \frac{\Lambda_{\rm QCD}^2}{F_a m_H} \bigg)^{0.3},
\eeq
where $m_H <m_a$ is assumed. 
See Fig.~\ref{fig:mass_evolve} for the typical evolution of the mass eigenvalues
as a function of the cosmic temperature $T$. We can see that $a_1$ and $a_2$ are initially identified as the QCD and hidden axion 
respectively for $T>T_{\rm res}$, and eventually they are exchanged with each other after the cross-over of the mass eigenvalues.

At  temperature $T \approx T_{\rm res}$,  QCD axions are converted to  hidden axions through resonance {\it a la} the MSW effect in 
neutrino physics \cite{Wolfenstein:1977ue}, and vice versa. This conversion process occurs efficiently if the adiabaticity condition is satisfied,
namely, if both axions oscillate many times over the Hubble time. Then, there is an adiabatic invariant
during the resonant conversion. In particular, if the axion potentials are approximated by the quadratic potential, 
it is the number of axions in a co-moving volume that is conserved, not the amplitude of oscillations. 
The hidden axion starts to oscillate when the Hubble parameter becomes comparable to its mass, $H \sim m_H$,  
unless it is initially located in the vicinity of the potential maximum. The oscillation amplitude of $a_H$ is likely so small 
 at the resonance
temperature  that its dynamics can be approximated by the harmonic oscillations.
To parametrize the adiabaticity, we define a parameter $\xi$ as a ratio of the Hubble parameter to the axion mass
evaluated at the resonance:
\beq
	\xi \equiv \frac{H(T_{\rm res})}{m_H} = \frac{H(T_{\rm res})}{m_a(T_{\rm res})} \simeq 4.4 \bigg(\frac{m_H}{m_a} \bigg)^{-1.6} \frac{F_a}{M_P}.
	\label{eq:adiabaticity}
\eeq
Note that the hidden axion mass is equal to the QCD axion mass at the cross-over temperature.
As we shall see shortly, the resonance occurs efficiently  for  $\xi \ll 1$, and becomes incomplete as $\xi$ approaches unity.
This sets the lower bound on $m_H$ for the efficient resonant conversion; this bound roughly reads $m_H \gtrsim H(T=\Lambda_{\rm QCD}) \simeq
2 \times 10^{-10}$\,eV.
The above adiabaticity parameter does not take account of the anharmonic effects, which become important when the
initial misalignment angle $\theta_i$ approaches $\pi$. 
We will see that the resonance becomes incomplete as we increase the initial misalignment angle of $a$. This can be understood
by noting that, when the anharmonic effect is important,  the typical time scale around the end points of oscillations is
 longer than the mass scale around the 
potential minimum, which enhances the breaking of the adiabaticity condition.

\subsection{Abundance}
\label{subsec:abundance}

The number of the QCD axions in a comoving volume is an adiabatic invariant that is conserved during the conversion processes,
if the adiabaticity parameter $\xi$ is much smaller than unity, and if the anharmonic effect is
 negligible. Therefore the resultant energy density of the axion CDM is expected to be suppressed by the mass ratio, $m_H/m_a$, compared to the
case without resonant conversion.

\begin{figure}[tp]
\centering
\subfigure[]{
\includegraphics [width = 7.5cm, clip]{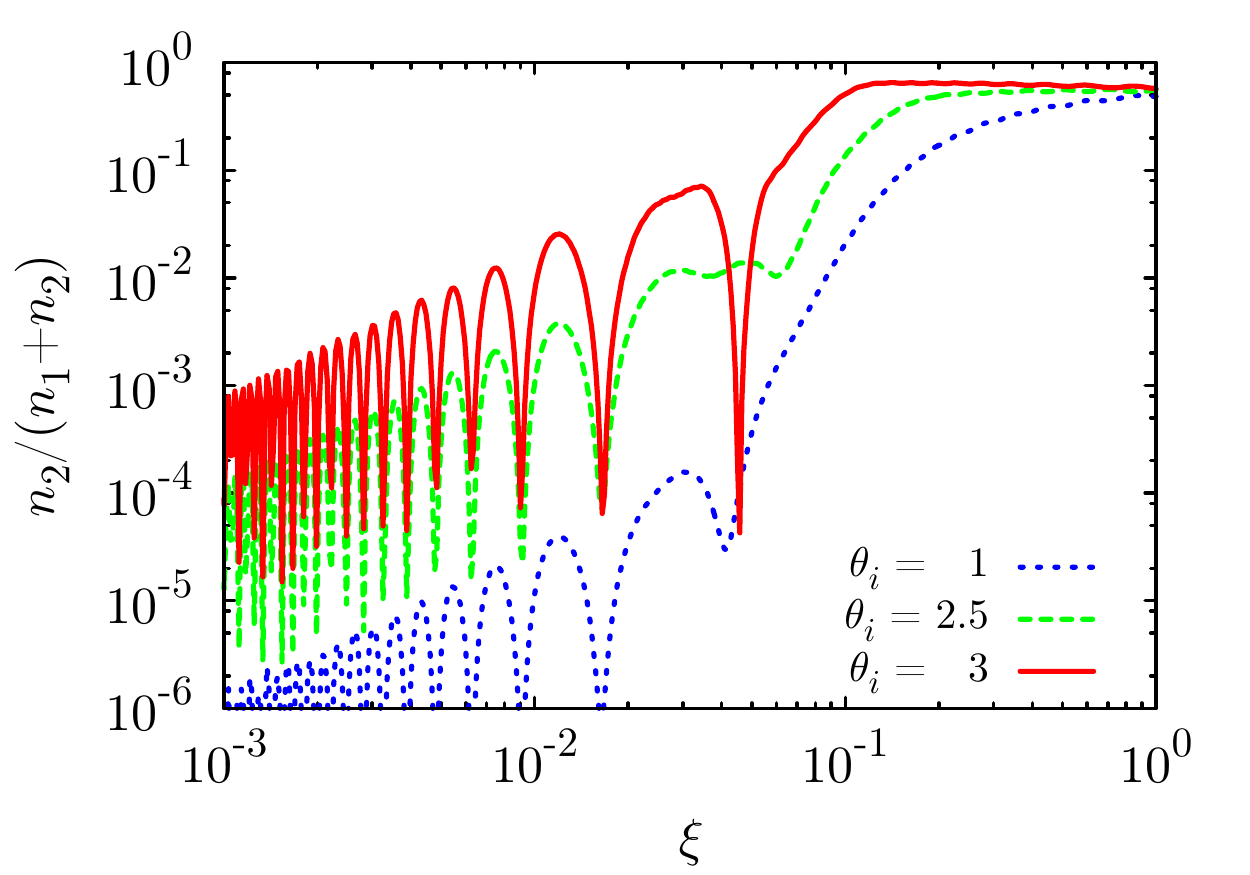}
\label{subfig:n-xi}
}
\subfigure[]{
\includegraphics [width = 7.5cm, clip]{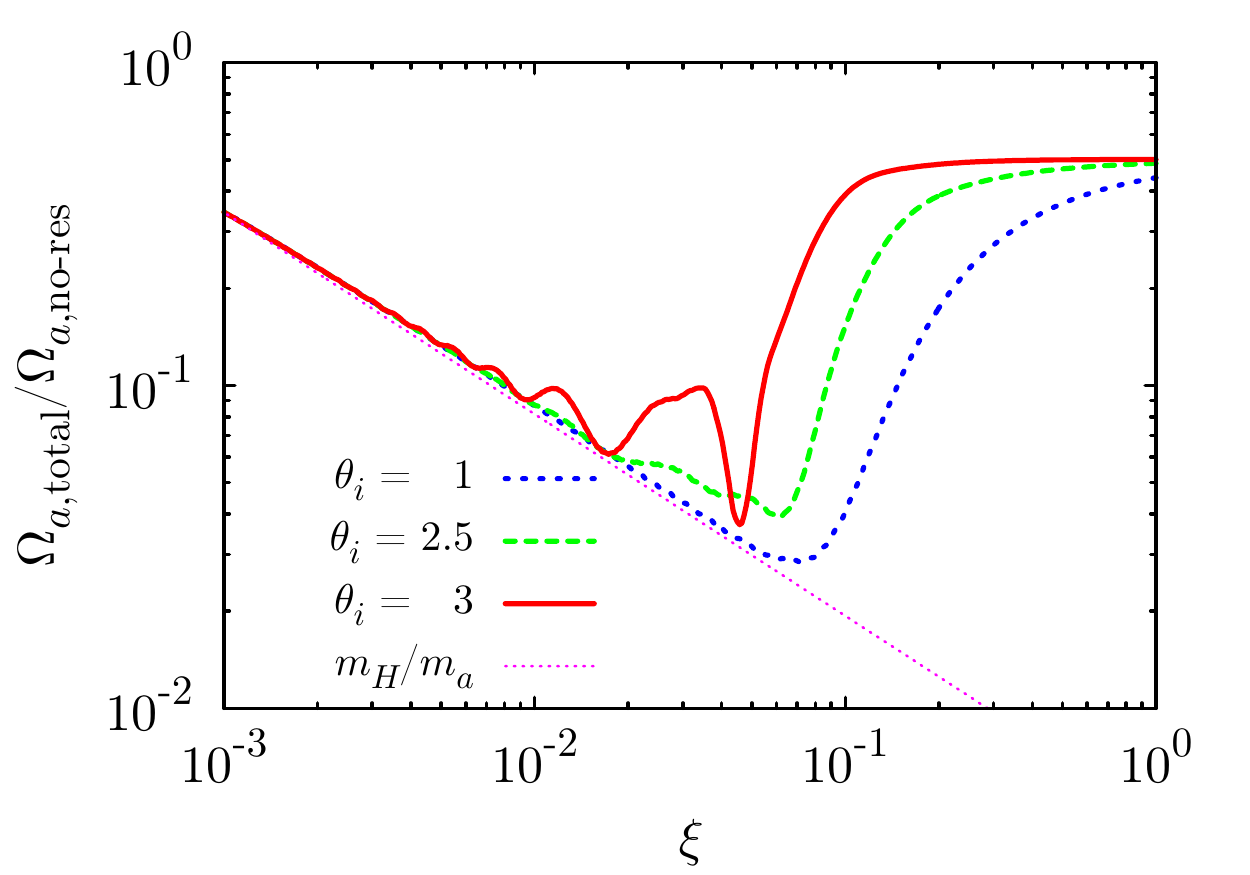}
\label{subfig:Omega-xi}
}
\caption{
	The ratio of the number density of $a_2$ to the total number density 
	 (left panel) and the suppression factor of the axion density parameter
	 with respect to the case without resonant conversion (right panel).
	The resonant conversion becomes efficient for small values of the adiabaticity parameter $\xi$.
	We have taken $F_a = F_H = 10^{14}~{\rm GeV}$, 
	and varied the initial misalignment angles as 
	 $\theta_i  \equiv a_i/F_a= 1$ (dotted blue),  $ 2.5$ (dashed green) and $ 3$ (solid red) with $\theta_{H,i} = -\theta_i$.
	The small dotted (magenta) straight line in the right panel shows $m_H/m_a$,
	and we can see that the resultant density parameter is indeed suppressed by the mass ratio for small $\xi \lesssim 0.1$.
}
\label{fig:supression}
\end{figure}

Too see this, 
we have numerically solved the axion dynamics with such initial condition that only coherent oscillations of $a_1$
are induced while $a_2$ initially sits at the potential minimum. The results are shown in
Fig~\ref{fig:supression}. Here we have taken $F_a = F_H = 10^{14}~{\rm GeV}$ 
and varied the initial misalignment angles as  $\theta_i  \equiv a_i/F_a= 1$ (dotted blue),  $ 2.5$ (dashed green) and
$ 3$ (solid red)
with $\theta_{H,i} \equiv a_{H,i}/F_H= -\theta_i$. 
With this initial condition, $a_1$ starts to oscillate when the Hubble parameter becomes comparable to
$m_a(T)$, while no coherent oscillations of $a_2$ are induced in the low energy, 
if the resonant conversion is $100$ percent efficiency. In the actual Universe, however, the adiabaticity condition is weakly
violated by a non-zero value of $\xi$, and as a result, a small amount of coherent oscillations of $a_2$ is induced.
In  Fig~\ref{subfig:n-xi}, we show the resultant number density ratio, $n_2/(n_1+n_2)$, as a function of the
adiabaticity parameter $\xi$ by varying $m_H$.
Here the number density of each axion is defined by the energy density divided by the mass.
One can see that only a tiny amount of $a_2$ is induced for small values of $\xi$, and the resonant conversion
occurs efficiently. On the other hand, 
a larger fraction of $a_1$ is converted to the heavier eigenstate $a_2$ as $\xi$ increases because of the
incomplete resonant conversion. Also, one can see that, as the initial misalignment angle $\theta_i$ increases,
more $a_2$ is induced due to the incomplete conversion. This is because the anharmonic effect tends to break the adiabaticity
condition, making the resonant conversion inefficient.

In Fig.~\ref{subfig:Omega-xi} we show the resultant density parameter of the total axion CDM, $\Omega_{a, {\rm total}} \equiv \Omega_{a1}+\Omega_{a2}$,
normalized by the QCD axion density parameter in the absence of the resonant conversion, $\Omega_{a,{\rm no-res}}$, 
as a function of $\xi$. 
As expected, the suppression factor is approximately given by the mass ratio,  $m_H/m_a$,  for small $\xi$, where
the resonant conversion is efficient. Note that we actually vary $m_H$ in this plot, while the other parameters are fixed. 
The suppression factor is of order $0.01$ for the parameters adopted; 
 we have chosen the decay constants that are slightly larger than the conventional axion window, since the numerical
computation would become expensive, otherwise. For smaller values of $F_a$ and $m_H$, the suppression factor $m_H/m_a$ becomes smaller, and so, 
the final axion abundance can be suppressed by a larger amount. 
Thus, we can reduce the axion CDM abundance by making use of the
resonant conversion of the QCD axions into hidden axions. Note that, although the initial misalignment angle for $a_2$ is set to be zero
in our numerical calculation, 
the total axion abundance can be still suppressed for a certain range of the misalignment angle, because
$a_2$ starts to oscillate before $a_1$ and its abundance tends to be suppressed compared to that of $a_1$. 
Interestingly, the suppression factor exhibits oscillating behavior for $\xi$, which sensitively depends on the initial misalignment angle.
This behavior affects the axion isocurvature perturbation as will be shown next.

\subsection{Isocurvature perturbations}
\label{subsec:isocurv}

If the PQ symmetry is already broken during the last $50$ or $60$ e-foldings of inflation, 
the QCD axion acquires quantum fluctuations, leading to
isocurvature perturbations imprinted on the CMB temperature anisotropy. Similarly, hidden axions also
give rise to isocurvature perturbations. 
The amount of isocurvature perturbations is tightly constrained by the recent CMB observations \cite{Ade:2013zuv}. 
Taking into account the anharmonic corrections, the CDM isocurvature perturbation from both the QCD and hidden axions is 
calculated as \cite{Kobayashi:2013nva}
\bea
	\Delta_{\mathcal{S},{\rm CDM}} &=& \bigg( \frac{\Omega_{a,{\rm total}}}{\Omega_{\rm CDM}}\bigg) \Delta_{\mathcal{S},a},
\eea
with\footnote{
Here we have truncated higher order terms, which would be important only when the leading term
is somehow suppressed. The effects of higher order terms are encoded in the
 non-Gaussianity of isocurvature perturbations, which we shall discuss later.
}
\bea	
	\Delta_{\mathcal{S},a} &=& \frac{\partial \ln \Omega_{a,{\rm total}}}{\partial \theta_i} \delta \theta_i 
	+ \frac{\partial \ln \Omega_{a,{\rm total}}}{\partial \theta_{H,i}} \delta \theta_{H,i},
\eea
where $\Omega_{\rm CDM}$ is the observed CDM density parameter, 
$H_{\rm inf}$  the inflationary Hubble parameter, and $\delta \theta_i$, $\delta \theta_{H,i}$ the quantum fluctuations of $\theta_i$, $\theta_{H,i}$
with
$\la \delta \theta_i^2 \ra = (H_{\rm inf}/2 \pi F_a)^2$, $\la \delta \theta_{H,i}^2 \ra = (H_{\rm inf}/2 \pi F_H)^2$.
Assuming that there is no correlation between $\delta \theta_i$ and $\delta\theta_{H,i}$, the power spectrum of the isocurvature perturbation is given by\footnote{
It is possible to modify the size of the quantum fluctuations if the radial component ``saxion" evolves during the last $50$ or $60$ 
e-foldings~\cite{Linde:1990yj,Kasuya:2009up}. A non-trivial correlation between $\delta \theta_i$ and $\delta \theta_H$ can be generated if some combination of the axions was
very heavy during inflation (cf. \cite{Jeong:2013xta}).
}
\beq
	\Delta^2_{\mathcal{S},a} = \bigg(\frac{\partial \ln \Omega_{a,{\rm total}}}{\partial \theta_i} \bigg)^2 \bigg( \frac{H_{\rm inf}}{2 \pi F_a} \bigg)^2
	+ \bigg( \frac{\partial \ln \Omega_{a,{\rm total}}}{\partial \theta_{H,i}} \bigg)^2 \bigg( \frac{H_{\rm inf}}{2\pi F_H} \bigg)^2.
\eeq
The current upper bound on the CDM isocurvature perturbation reads~\cite{Ade:2013zuv}
\beq
\beta \;<\;0.039~~(95\%{\rm \,C.L.}),
\eeq
where $\beta$ is defined by
\beq
	\Delta^2_{\mathcal{S},{\rm CDM}} = \frac{\beta}{1-\beta} \Delta^2_\mathcal{R}.
\eeq
with $\Delta^2_\mathcal{R} \approx 2.2\times 10^{-9}$ being the curvature power spectrum.
This sets stringent constraints on the inflation scale, which were studied in the literature in the case of only QCD 
axions~\cite{Higaki:2014ooa,Marsh:2014qoa,Visinelli:2014twa}.

In the absence of the resonant conversion, the abundance of QCD axions is a monotonically increasing function of 
the initial misalignment angle $\theta_i$. In particular, it rapidly increases as $\theta_i$ approaches $\pi$,
and so,  $\Delta_{\mathcal{S},a}^2$ is also an increasing function of $\theta_i$.
This is the reason why the isocurvature perturbations get enhanced toward the hilltop initial 
condition~\cite{Lyth:1991ub,Kobayashi:2013nva}.\footnote{In addition, the non-Gaussianity is  enhanced in the
hilltop limit.~\cite{Kobayashi:2013nva}.
\label{ftnt}
}

\begin{figure}[tp]
\centering
\includegraphics [width = 14cm, clip]{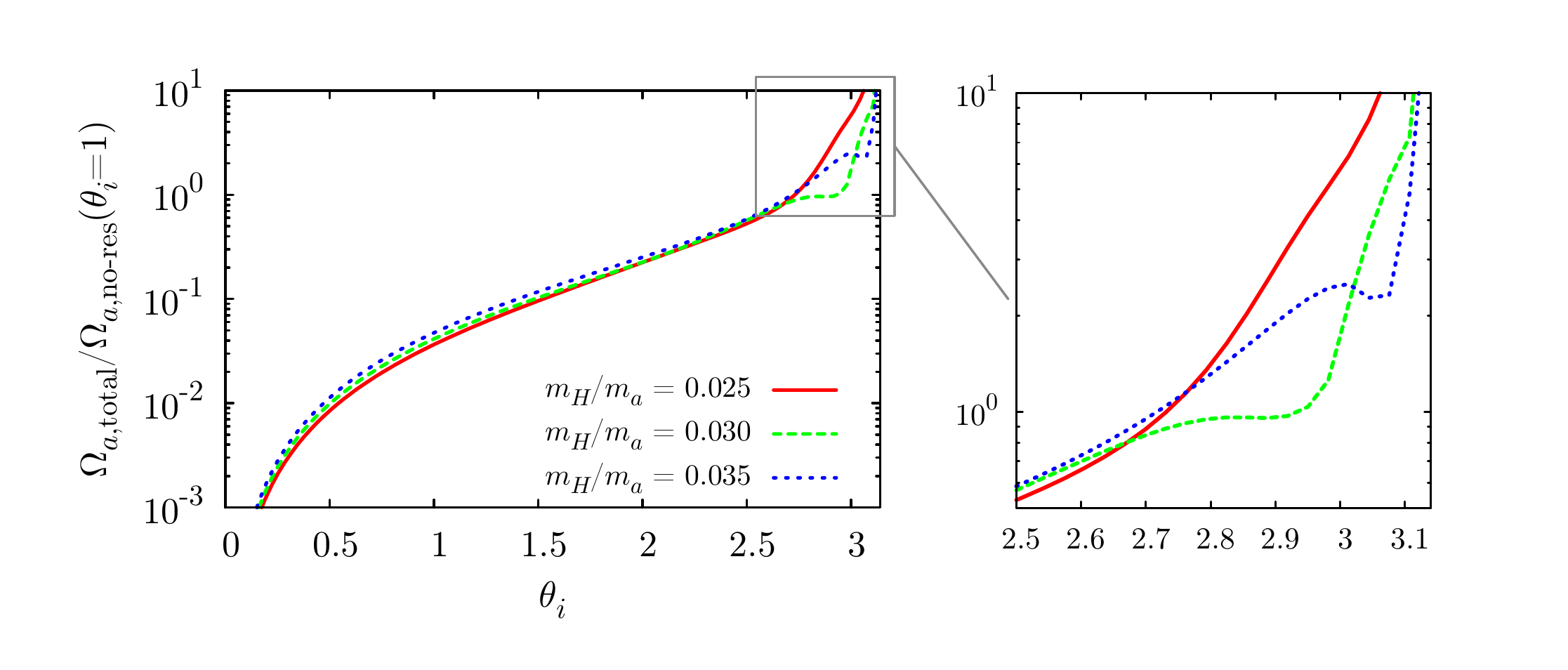}
\caption{
	The $\theta_i$-dependence on the density parameter of the axion CDM is shown.
	The vertical axis is normalized by the density parameter in the single QCD axion case with $\theta_i = 1$.
	We have taken $F_a = F_H = 10^{14}~{\rm GeV}$, $\theta_{H,i} = -\theta_i$  
	and $m_H/m_a = 0.025$ (solid red), 0.03 (dashed green), 0.035 (dotted blue).
}
\label{fig:Omega-theta}
\end{figure}

\begin{figure}[tp]
\centering
\includegraphics [width = 15cm, clip]{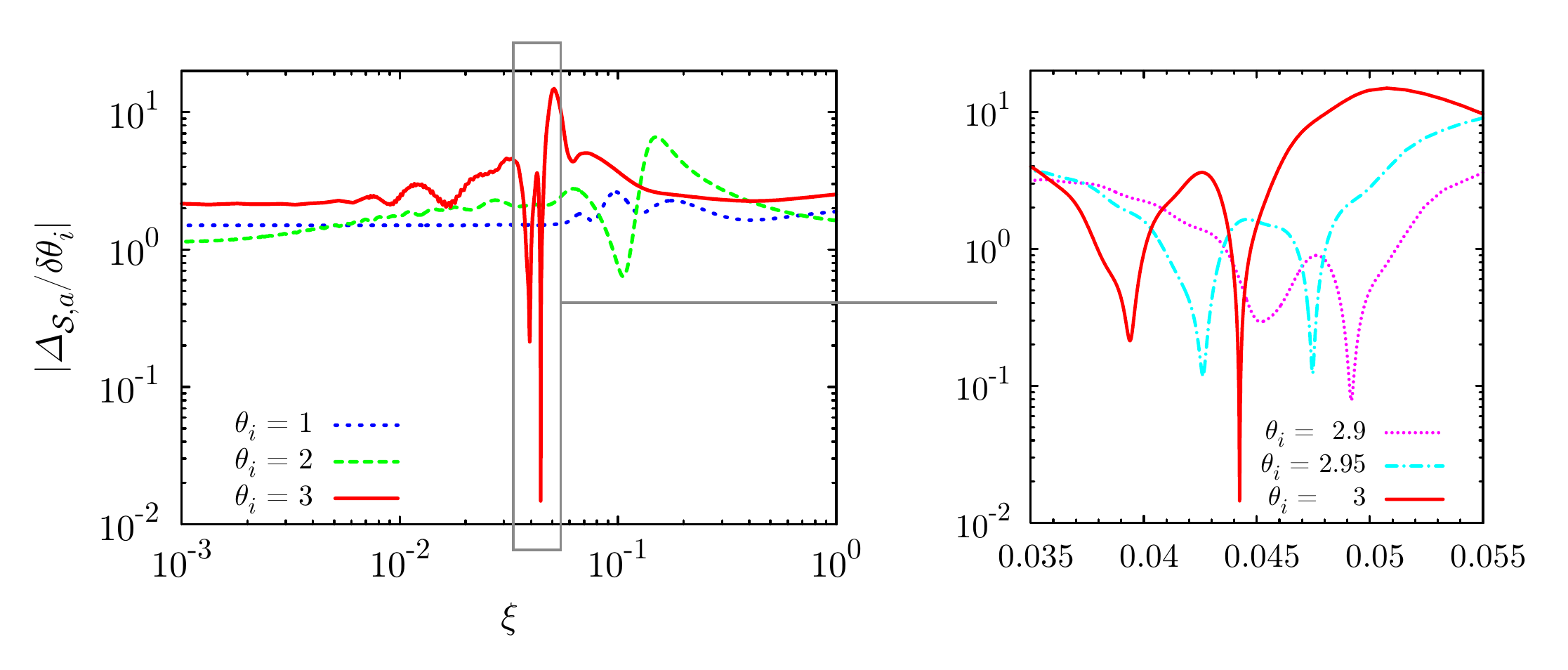}
\caption{
	The normalized isocurvature perturbation of the axion CDM,
	$|\Delta_{\mathcal{S},a}/\delta\theta_i|$ (see Eq.~(\ref{eq:dsdt})),
	is shown as a function of  the adiabaticity parameter defined in Eq.~(\ref{eq:adiabaticity})
	for several values of the initial misalignment angle $\theta_i$.
	We have taken $F_a = F_H = 10^{14}~{\rm GeV}$, 
	$\theta_{H,i} = -\theta_i$ 
	and $\theta_i = 1$ (dotted blue), 2.5 (dashed green), 3 (solid red), 
	$2.9$ (small-dotted magenta) and 2.95 (dash-dotted cyan).
}
\label{fig:isocurv}
\end{figure}

If there is a resonant conversion of  QCD axions into hidden axions, the situation is different.
This is because the conversion rate depends on both $\theta_i$ and the mass ratio $m_H/m_a$.
In Fig~\ref{fig:Omega-theta}, we show  $\Omega_{a, {\rm total}}$ as a function of $\theta_i$ for various
values of $m_H/m_a$.  One can see that there is a plateau around $\theta_i \simeq 3$ for $m_H/m_a = 0.03$,
where the isocurvature perturbations are expected to be significantly suppressed.\footnote{Note that the axion abundance exceeds the 
observed dark matter for the adopted parameters, because  $F_a = \GEV{14}$ is chosen for efficient numerical
calculation. The purpose of this letter is to study the effect of the resonant
conversion on the isocurvature perturbations, and we leave the calculation with smaller values of $F_a$ for future work, as the required numerical
computation is more expensive. Similar suppression is expected for the case with smaller values of $F_a$. }
In Fig.~\ref{fig:isocurv}, we show the axion isocurvature perturbations 
normalized by the fluctuation of the misalignment angle of the QCD axion, 
\beq
\left|\frac{\Delta_{\mathcal{S},a} }{\delta\theta_i}\right| = \sqrt{\lrfp{\partial\ln\Omega_{a,{\rm total}}}{\partial \theta_i}{2}
+\lrfp{F_a}{F_H}{2} \lrfp{\partial \ln\Omega_{a,{\rm total}} }{ \partial \theta_{H,i}}{2}},
\label{eq:dsdt}
\eeq
as a function of the adiabaticity parameter. 
One can see that there is indeed a significant suppression
of the isocurvature perturbations at specific values of $\xi$ and $\theta_i$. 
The detailed structure of the suppression is shown in the right panel of Fig.~\ref{fig:isocurv}; 
there are points where isocurvature perturbations are significantly suppressed. In the vicinity of these points
$\partial \Omega_{a,{\rm total}}/\partial \theta_i$ vanish and its sign flips between these points.

\section{Discussion and conclusions}
\label{sec:conc}

In this letter, we have studied the model of the QCD and hidden axions with a mass mixing 
and performed numerical calculations to follow the MSW-like conversion process between them,  taking account of
a weak violation of non-adiabaticity as well as the anharmonic effects.
To characterize the violation of the adiabaticity, we have introduced an adiabaticity parameter, $\xi$, defined by
the ratio of the Hubble parameter to the axion mass at the resonance (see Eq.~(\ref{eq:adiabaticity})), 
and we have found that the resultant axion CDM abundance can be suppressed by a factor of $m_H/m_a$ if the
resonant conversion is efficient, i.e., if the adiabaticity parameter is much smaller than unity and the anharmonic
effects are negligible. 
Furthermore, we have found that the anharmonic effect makes the resonant conversion less efficient and, 
most interestingly, it significantly  affects the axion CDM isocurvature perturbations. 
We have shown that the axion CDM isocurvature perturbations can be suppressed 
for certain values of the parameters where the resonant conversion is incomplete. 
In the following we mention on the limitations and implications of the present analysis.

 We have performed the calculation with the initial condition that the hidden axion direction sits at the minimum, 
 $\theta_i + \theta_{H,i} = 0$,  in order to focus on the resonant conversion from  QCD axions
 to hidden axions. For a more general initial condition, the resonant conversion from hidden axions to QCD axions also
 takes place, which complicates the dynamics of axions. In fact we have calculated such cases and confirmed
 that the isocurvature perturbations can be similarly suppressed for a certain range of the parameters.
 In order to estimate the suppression factor quantitatively, we need to scan the parameter space
in the $\theta_i$~--~$\theta_{H,i}$ plane. 
Note that $m_H$ needs to be much smaller than $m_a$ to suppress the axion abundance, 
but the commencement of the hidden axion oscillations is delayed for small $m_H$ and the amplitude may not be damped sufficiently at the resonance.
Thus, there is a limitation to the suppression and we will investigate  the conditions under which the total axion CDM
abundance is maximally reduced.

Throughout the letter, we have taken $F_H = F_a$ for simplicity.
If we choose a smaller value of $F_a$ with $F_a< F_H$, the QCD axion mass at the zero temperature is increased, and as a result, the total axion DM
abundance will decrease as it is suppressed by a factor of $m_H/m_a$.
On the other hand, if we choose $F_a > F_H$, the hidden axion oscillations induced by the resonant conversion of the QCD axion
can climb over the cosine potential hill 
and roll down to the adjacent minimum, $a_{H,{\rm min}} = 2\pi F_H$.
We have confirmed this behavior numerically.
This implies that domain walls may be formed by the resonant conversion.
Once domain walls are formed and if they are stable, they will dominate the Universe soon and our present Universe cannot be realized.
We will also study the parameter region to avoid such a domain wall formation in the future.

So far we have considered only a Gaussian part of the isocurvature perturbations, but higher-order
terms become important when the leading Gaussian part is suppressed.
The constraint on non-Gaussianity of the isocurvature perturbations is characterized by the parameter, $f_{\rm NL}^{({\rm iso})}$ 
and the current constraint reads \cite{Hikage:2012be}
\beq
	\bigg(\frac{\beta}{1-\beta} \bigg) f_{\rm NL}^{({\rm iso})} = 40 \pm 66.
\eeq
While the exact form of  $f_{\rm NL}^{({\rm iso})}$ in the multi-axion case is rather involved, it is roughly estimated as 
$\beta f_{\rm NL}^{({\rm iso})} \sim \partial^2 \ln \Omega_{a,{\rm total}}/\partial \theta_i^2,~\partial^2 \ln \Omega_{a,{\rm total}}/\partial \theta_{H,i}^2$.
We have also checked that the non-Gaussianity of the isocurvature perturbations are not suppressed in general, even if
the Gaussian part is suppressed due to the incomplete resonant conversion.\footnote{
Vice versa; we observe  that the non-Gaussianity can be suppressed for a certain range of the parameters, where
the Gaussian part is not suppressed.
}
These second derivatives are of $\mathcal{O}(10-100)$ for the parameters of our interest, and
the current constraints from non-Gaussianity of the isocurvature perturbation can be (marginally) satisfied.
The mild enhancement of the non-Gaussianity is a necessary outcome of our scenario, because the anharmonic
effect plays a crucial role in suppressing the power spectrum of the isocurvature perturbations (cf. footnote \ref{ftnt}).
It is worth studying how the resonant conversion affects non-Gaussianity, which we leave for future work.

We have focused on the resonant conversion between the QCD and hidden axions so far. In principle a similar resonant conversion
could take place between two hidden axions with a mass mixing\footnote{
This argument is not limited to axions, but, in principle, it can be applied to coherent oscillations of any scalars. 
}, if one of them gradually acquires a mass and there appears a 
cross-over of the mass eigenvalues. The isocurvature perturbations of the hidden axions can be suppressed similarly for a certain
set of the parameters. 

\section*{Acknowledgment}
This work was supported by  JSPS Grant-in-Aid for
Young Scientists (B) (No.24740135 [FT]), 
Scientific Research (A) (No.26247042 [FT]), Scientific Research (B) (No.26287039 [FT]), 
 the Grant-in-Aid for Scientific Research on Innovative Areas (No.23104008 [NK, FT]),  and
Inoue Foundation for Science [FT].  This work was also
supported by World Premier International Center Initiative (WPI Program), MEXT, Japan [FT].



\end{document}